\documentclass[twocolumn,showpacs,preprintnumbers,amsmath,amssymb]{revtex4}

\usepackage{graphicx,color}
\usepackage{dcolumn}
\usepackage{bm}
\usepackage{CJK}

\begin{document}

\title{Relativistic nucleon optical potentials with isospin
dependence in Dirac Brueckner Hartree-Fock approach}

\author{Ruirui Xu }
\author{Zhongyu Ma}
\affiliation{China Institute of Atomic Energy, P.O. Box 275(41), Beijing 102413, China}%
\author{E. N. E. van Dalen}
\author{H. M\"{u}ther}
\affiliation{Institut f\"{u}r Theoretische Physik, Universit\"{a}t T\"{u}bingen,
Auf der Morgenstelle 14, D-72076 T\"{u}bingen, Germany}%

\date{\today}

\begin{abstract}

The relativistic optical model potential (OMP) for nucleon-nucleus scattering
is investigated in the framework of Dirac-Brueckner-Hartree-Fock (DBHF) approach using the Bonn-B
One-Boson-Exchange potential for the bare
nucleon-nucleon  interaction. Both real and imaginary parts of isospin-dependent
nucleon self-energies in nuclear medium are  derived from the DBHF approach
based on the projection techniques within the subtracted $T$-matrix
representation. The Dirac potentials as well as the corresponding
Schr\"{o}dinger equivalent potentials are evaluated.
An improved local density approximation  is
employed in this analysis, where a range parameter is included to account for a
finite-range correction of the nucleon-nucleon interaction. As an example the
total cross sections, differential elastic scattering cross sections, analyzing
powers for $n$, $p$ + $^{27}$Al at incident energy 100 keV $\leqslant~ E
~\leqslant$ 250 MeV are calculated. The results derived from this microscopic
approach of the OMP are compared to the experimental data, as well as the
results obtained with a phenomenological OMP. A good agreement between the
theoretical results and the measurements can be achieved for all incident
energies using a constant value for the range parameter.
\end{abstract}

\maketitle

\section{Introduction}

The Optical Model potential (OMP) is one of the essential tools of
exploring the physics of nuclear reactions. The investigations of OMP
based on a fundamental microscopic theory attracts a lot of
interest because it provides a reliable basis to explore
nuclear reaction mechanisms in particular for cases without
experimental data. With the development of radioactive
beam facilities, many new physical phenomena have been and will be observed for exotic
nuclei and unstable isotopes beyond the $\beta$ stable line. A
number of nuclei with large neutron excess will be produced and
their structure and collisions are under investigation. Thus, the
isospin dependence of the optical potentials  becomes an important
issue in the systematic study of OMP. However, this feature was
underdeveloped for lack of direct guidance from measurements in
the past. Therefore, the main purpose of this work is to construct
the nucleon OMP in a microscopic way and discuss its
features, especially with respect to the isospin
dependence.

Different models have been used to determine a microscopic OMP
\cite{Jeu77,Ma88,Bonats90,Li93,CM.95,Ron06,LiZ08,Han10}. They can be
categorized as semi-phenomenological and {\it ab initio} approaches
according to the nucleon-nucleon
interaction and the approximation scheme used to describe the many-body systems.
The semi-phenomenological approaches normally begin with a
phenomenological nucleon-nucleon ($NN$) effective interaction treated within a
mean field approximation. Typical examples are the Skyrme-Hartree-Fock
(SHF)\cite{Han10} and
the relativistic mean-field (RMF)\cite{Ma88}. In these approaches, the
$NN$ effective interactions are phenomenologically adjusted to reproduce
nuclear matter saturation properties and the ground state properties
of a few well known stable nuclei.
The imaginary parts of
OMP are determined in an independent manner through the evaluation of the
polarization diagram. On the other hand, the {\it ab initio} method, such as the
Brueckner-Hartree-Fock (BHF)\cite{Jeu77,Bonats90,LiZ08,Zuo06} and
Dirac-Brueckner-Hartree-Fock (DBHF)
\cite{Bou87,BM.90,Sammar07,CM.95,Dalen04,Dalen07,Dalen10} approaches,
starts from a microscopic model of the bare $NN$
interaction and the nuclear many-body problem is
handled in a more sophisticated manner.
Since the $NN$ interaction is adjusted to describe the two-nucleon data,
the many-body calculation does not contain any adjustable parameter. Therefore such studies
should have a larger predictive power when applied in the study of exotic nuclear systems.

It is well known that the non-relativistic BHF
fails to reproduce  the empirical saturation properties of nuclear
matter without an extra three-body force\cite{LiZ08,Zuo06}. An
encouraging success has been obtained when the problem is addressed
in a relativistic framework, namely the DBHF
approach\cite{Bou87,BM.90,Sammar07,Muther01}. The saturation properties of
nuclear matter in DBHF are described rather accurately without any need to add
a three-body force. Moreover, another important feature of the
relativistic approach applied to finite nuclei is, that the spin-orbit potential arises
naturally from the coherent sum of the contribution from the scalar
and vector potentials \cite{Ron06}, which implies that this important ingredient
of OMP occurs in the DBHF treatment without the need of any adjustments.

The optical potential of a nucleon in the nuclear medium corresponds to the nucleon
self-energy \cite{Bell59}.  In the DBHF approach one can distinguish between two
frequently used schemes~\cite{Dalen10b} to determine the relativistic components of
the self-energy:  the fit method versus the projection technique method. In the fit
method, originally proposed by Brockmann and Machleidt~\cite{BM.90}, the scalar and
vector self-energy components are directly extracted from the single particle
energy. A fit to describe the momentum dependence of the single-particle energy in
terms of constant scalar and  vector components defines these  self-energy
components for each density considered.   An attempt, which tries to extend this
method in order to extract momentum dependent scalar and vector components  by
elaborate fitting procedures~\cite{lee97}, suffers from large uncertainties.
Therefore, only mean values for the self-energy components, where the explicit
momentum dependence has already been averaged out, can be obtained.  In symmetric
nuclear matter this method turned out to be rather reliable and predicts a
relativistic decomposition of the self-energy in close agreement to the analysis of
the projection technique. The extension of this simple method to the  isospin
asymmetric nuclear matter, however, fails to determine the correct  behavior of the
isospin dependence of the nucleon self-energies~\cite{Muther01,UM.97}.  Therefore,
this method is not very well suited for the construction of the microscopic OMP.

In the other method, the projection technique method, the scalar and vector
components of the self-energies are directly determined from the projection onto
Lorentz invariant amplitudes. These projection techniques are rather involved but
more accurate.  It requires the knowledge of the Lorentz structure of the
positive-energy-projected in-medium on-shell $T$-matrix. However,
ambiguities~\cite{nuppenau89}  arise due to the restriction to positive energy
states, since pseudo-scalar ($ps$) and pseudo-vector ($pv$) components can not
uniquely be disentangled for on-shell scattering. As a consequence an unphysical
strong momentum dependence of the nucleon self-energy is obtained, when using the
$ps$ representation scheme~\cite{Gross99}. This strong momentum  dependence mainly
originates from the contribution of the single $\pi$ exchange. Therefore, it was
suggested to separate the single meson exchange contributions, i.e. the bare interaction
$V$, and the high-order corrections $\Delta T$~\cite{Gross99}. The contributions of
single $\pi$ and $\eta$ exchange are in this approach  treated with the complete
pseudo-vector (pv) representation, whereas the pseudo-scalar (ps) representation is
applied only on the remaining part of  $T$-matrix, i.e. the high-order corrections
and the single meson exchanges of the other mesons. This representation scheme for
the $T$-matrix is called the subtracted $T$-matrix (STM). The application of this
representation scheme to the single meson exchanges reproduces the Hartree-Fock
nucleon self-energies.  Therefore, it is in principle equivalent to the method used
in the work of Ref.~\cite{Muther01}, which not only considered symmetric nuclear
matter but also pure neutron matter. This method was successfully applied to the
investigation of the microscopic OMP\cite{Ron06} as well as the nucleus-nucleus
scattering with a double folding method\cite{ZTM.08}.

Recently an obvious improvement has been achieved in the DBHF calculations using
projection techniques by the extension to the general case of isospin  asymmetric
nuclear matter, in which the neutrons and protons are occupying different Fermi
spheres~\cite{Dalen04,Dalen07,Dalen10}. As a consequence, one has different
effective masses and self-energies for neutrons and protons. Furthermore, one has to
deal with three different Pauli operators and in-medium interactions of the
nucleons, to be specific the $nn$, $pp$, and $np$ ones.  In contrast to the five
independent helicity matrix elements  for identical particles,  six helicity matrix
elements are independent in the $np$ channel. Hence, instead of the five Lorentz
invariants in the $nn$ and $pp$ channel, the $T$-matrix for the $np$ channel is
projected onto six Lorentz invariants. With this method one can obtain the real part
and the imaginary part  of the self-energy for symmetric and asymmetric nuclear
matter.

This method has been applied in this work to investigate
the isospin-dependent relativistic microscopical optical potential (RMOP).
To obtain the OMP for
finite nuclei, an improved local density approximation (ILDA) method
\cite{Jeu77} is employed to determine the  spacial distributions of
the corresponding self-energies from the relevant
density distribution and asymmetry of the finite nucleus under consideration.
As an example we analyze the
elastic scattering reactions of $n, p +^{27}$Al in terms of the RMOP
and compare with experimental data and those obtained from an analysis with
a phenomenological OMP.

The paper is arranged as follows. The formalism of DBHF in STM
representation is briefly introduced and the nuclear matter
properties with Bonn-B are described in Sec. II. In Sec. III we
present the isospin-dependent RMOP  and some relevant discussions.
The applications of  RMOP to the neutron and proton elastic
scattering off $^{27}$Al are shown in Sec. IV. Finally, a brief
summary is presented in Sec. V.

\section{DBHF APPROACH IN ASYMMETRIC NUCLEAR MATTER}

\subsection{ DBHF approach }

In the relativistic DBHF scheme, the interaction
between nucleons in nuclear matter is determined
through the ladder approximation of the relativistic Bethe-Salpeter (BS)
equation,

\begin{eqnarray}
     T = V + i \int VQGGT, \label{eq1}
\end{eqnarray}
where $T$ represents the nucleon-nucleon interaction matrix in the
nuclear medium and $V$ is the bare $NN$ interaction, respectively.
The Pauli exclusion principle is included by $Q$ operator and the
in-medium nucleon propagation is introduced by the Green's function
$G$, which fulfills the Dyson equation,

\begin{eqnarray}
     G = G_0 + G_0 \Sigma G. \label{eq2}
\end{eqnarray}
$G_{0}$ denotes the free nucleon propagator, and the self-energy
term $\Sigma$ can easily be derived in first order through the
following standard treatment

\begin{eqnarray}
     \Sigma = - i \int_F (Tr[GT]-GT), \label{eq3}
\end{eqnarray}
in which the self-energy contains the direct and exchange forms at
the same time, and the momentum is integrated within the Fermi sea.
Because the Eqs. (\ref{eq1})-(\ref{eq3}) are strongly coupled, they
have to be solved iteratively until the
convergence is reached.

A one boson exchange potential (OBEP) model of the bare $NN$ interaction is
normally employed for the $V$ term in Eq. (\ref{eq1}), and Bonn-B $NN$
interaction is selected in this discussion. Six mesons with
different (J$^\pi$,T) are involved in the Bonn-B interaction,
including two scalars $\sigma$~(0$^+$,0) and $\delta$~(0$^+$,1), two
vectors $\omega$~(1$^-$,0) and $\rho$~(1$^-$,1), and two
pseudo-vectors (pv) $\eta$~(0$^-$,0) and $\pi$~(0$^-$,1).
For the description of isospin asymmetric nuclear matter, one needs to distinguish
between protons and neutrons. Therefore, this Bonn-potential code has been made
suitable to treat distinct particles in the medium~\cite{Dalen07}. As a consequence,
one has three different in-medium interactions of the nucleons, to be specific the
$nn$, $pp$, and $np$ ones. The $np$ channel has an additional independent amplitude
compared to the five independent amplitudes in the $nn$ and $pp$ channel.

The subtracted $T$-matrix (STM)
representation has been applied for the projection of the $T$-matrix in the BS equation.
The contributions of single $\pi$ and $\eta$ are in this approach
treated with the complete pv representation.  The ps representation is
applied only on the remaining part of the  $T$-matrix,

\begin{eqnarray}
     T_{sub} = T - V_{\pi,\eta}~.  \label{eq4}
\end{eqnarray}
In this STM representation the ambiguity problem of the self-energy from the
contributions of $\pi$ and $\eta$ mesons in isospin asymmetric nuclear
matter can be minimized satisfactorily.
Thorough discussions on generating the nucleon self-energies and comparing the
reproduced bulk properties of the nuclear matter with the empirical values for the
symmetric and asymmetric matter can be found in Ref.
\cite{Gross99,Dalen04,Dalen07,Dalen10}.

The relativistic nucleon self-energy in the asymmetric
nuclear matter is defined to satisfy the translational invariance
of the homogeneous system, and its common Lorentz structure
can be expressed as,

\begin{eqnarray}
     \Sigma^t(k,k_F,\beta) = \Sigma^t_s(k,k_F,\beta) - \gamma_0\Sigma_0^t(k,k_F,\beta) \label{eq5}\\
     \nonumber + \bm{\gamma}\cdot
     \textbf{k}\Sigma_v^t(k,k_F,\beta)~.
\end{eqnarray}
where $\Sigma_s$ is the scalar part of self-energy, $\Sigma_0$ and
$\Sigma_v$ denote the timelike and spacelike terms of the vector part,
respectively. The superscript $t$ is used to mark the proton and
neutron, since one has to distinguish between neutrons and protons in isospin asymmetric nuclear matter.
 It can be seen that the self-energies are the functions of
the nucleon momentum ($k$), density or Fermi momentum ($k_F$), and
asymmetry parameter $\beta$ = ($\rho_n$-$\rho_p$)/$\rho$, where
$\rho_n$, $\rho_p$ and $\rho$ indicate the neutron, proton and total
densities in nuclear matter.

\subsection{Nucleon self-energies and nuclear matter properties }
\begin{figure}[htbp]
\centerline{\includegraphics[width = 3.2in]{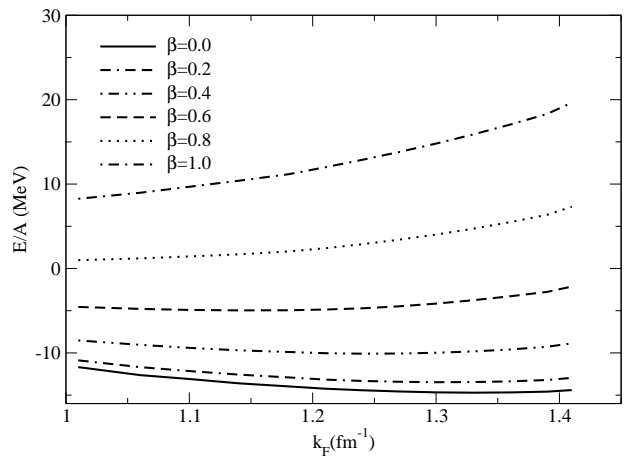}} \caption{
Binding energy per nucleon in
 nuclear matter.} \label{figure1}
\end{figure}

In this subsection, we present the bulk properties of nuclear matter
as well as the nucleon self-energies
in the DBHF approach with Bonn-B $NN$ interaction calculated according to
the procedure described in Ref.~\cite{Dalen10}.
The binding energies per nucleon (E/A) are shown in Fig. \ref{figure1} as a
function of the density, to be precise as a function of the corresponding
Fermi momentum in symmetric nuclear matter. Various values for the asymmetry
parameter $\beta$ are considered in the range from 0.0 (symmetric matter) to
1.0 (neutron matter). The energy versus density curves do not exhibit a minimum
for $\beta \geq 0.8$. The saturation point calculated in symmetric nuclear matter
occurs at $k_F$ = 1.33 fm$^{-1}$ with  E/A
= -14.71 MeV. This is  in good agreement with the empirical values
 $k_F$ = 1.36 $\pm$ 0.06 fm$^{-1}$ with E/A
= -16 $\pm$ 1 MeV. Furthermore this figure demonstrates, that, at a given density,
isospin asymmetric nuclear matter gets
monotonically less attractive with increasing $\beta$.

\begin{figure}[htbp]
\centerline{\includegraphics[width = 3.2in]{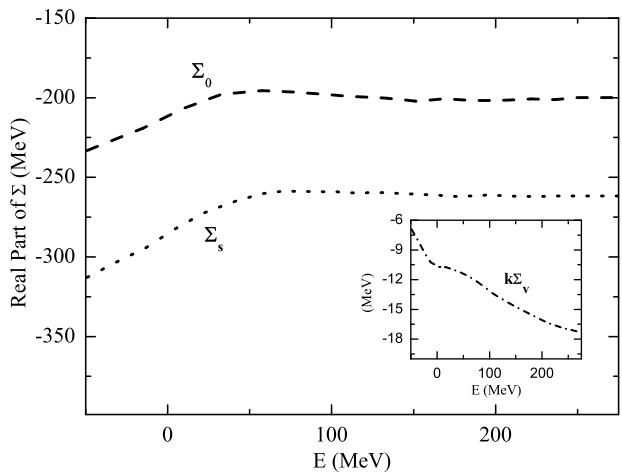}} \caption{Real
parts of the nucleon self-energies at $k_F$=1.35 fm$^{-1}$ in
symmetric nuclear matter. Note that $k$ is the momentum of single
particle.} \label{figure2}
\end{figure}

\begin{figure}[htbp]
\centerline{\includegraphics[width = 3.2in]{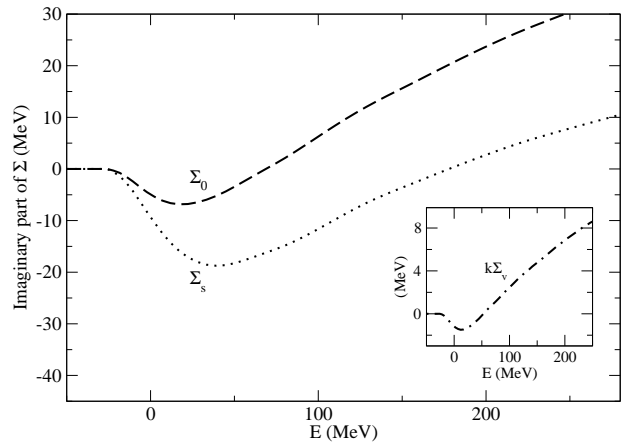}} \caption{Same
as Fig.\ref{figure2}, except for the imaginary parts.}
\label{figure3}
\end{figure}

For the example of symmetric nuclear matter at a Fermi momentum $k_F$=1.35 fm$^{-1}$
the real and imaginary parts of the nucleon self-energy components
$\Sigma_s$, $\Sigma_0$, and $\Sigma_v$ are shown in Fig. \ref{figure2} and Fig.
\ref{figure3}, respectively, as functions of the on-shell
nucleon energy $E$.
This single particle energy $E$ is
determined by
\begin{eqnarray}
      E=\sqrt{\textit{k}^2(1+\Sigma_v)^2+(M+\Sigma_s)^2}-\Sigma_0 - M  \label{eq12}.
\end{eqnarray}
In these figures, one can observe that the real parts of $\Sigma_s$
and $\Sigma_0$ change weakly with $E$ in Fig. \ref{figure2}
especially for the energies above 100 MeV.  A stronger energy
dependence is obtained for the imaginary parts as can be seen in Fig. \ref{figure3}.
Note that the imaginary parts vanish at the energy $E$ which corresponds to the
Fermi energy.
The real spacelike part $\textbf{\textit{k}}\Sigma_v$ is rather
small in comparison to the other two components, while the imaginary
$\textbf{\textit{k}}\Sigma_v$ turns out to be comparable in size with
Im$\Sigma_s$, Im$\Sigma_0$.

The equivalent potential to be used in a Schr\"odinger equation may in a first
approximation (more detailed discussion see below) be represented by the difference
$\Sigma_s-\Sigma_0$. Hence the real parts of the self-energy components compensate
each other to a large extent leading to a constant value for $\Sigma_s-\Sigma_0$ of
approximately -60 MeV for large energies $E$. This kind of cancellation between
scalar and vector components is not observed for the imaginary part. Nevertheless
also the imaginary part of $\Sigma_s-\Sigma_0$ is essentially constant around -20
MeV at larger energies.

\begin{figure*}[htbp]
\centerline{\includegraphics[width = 6in]{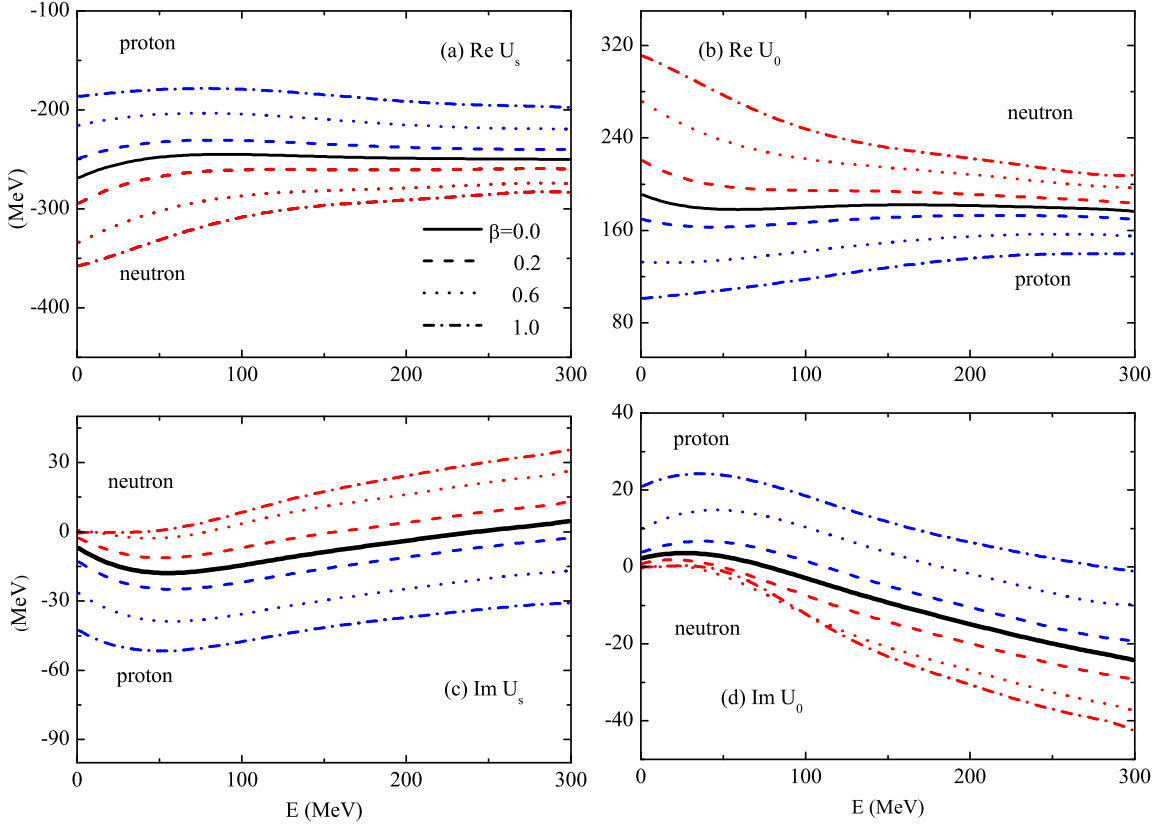}} \caption{(Color
online) Comparison of $U_s$,$U_0$ at $k_F$=1.35 fm$^{-1}$. The real
parts (upper panels) and imaginary parts (lower panels) are plotted
for various $\beta$ values. The black curves are the results for the
symmetric nuclear matter, the blue curves denote those for proton
and red curves for neutron. } \label{figure4}
\end{figure*}

\section{RELATIVISTIC MICROSCOPIC OPTICAL POTENTIAL IN FINITE NUCLEI}

The Dirac equation of a nucleon in the nuclear medium can be
written
\begin{eqnarray}
      \left[\vec{\alpha}\cdot\vec{p}+\gamma_0(\textit{M}+U_s^t)+U_0^t \right]
      \Psi^t=\varepsilon\Psi^t~ , \label{eq10}
\end{eqnarray}
where $U_s^t$ and $U_0^t$ are scalar and vector potentials,
\begin{eqnarray}
      U_s^t=\frac{\Sigma_s^t-\Sigma_v^t\textit{M}}{1+\Sigma_v^t},~~U_0^t=\frac{-\Sigma_0^t+\varepsilon\Sigma_v^t}{1+\Sigma_v^t}.~~ \label{eq11}
\end{eqnarray}
and $\varepsilon$ = $\textit{E}+\textit{M}$ is the single particle
energy, $E$ is the kinetic energy of the nucleon in the free space
and $M$ indicates the mass of the nucleon.

The Dirac potentials $U_s$ and $U_0$ with various values for the isospin
asymmetries $\beta=$ 0.0, 0.2, 0.6, 1.0 at $k_F=$1.35 fm$^{-1}$ are
plotted as functions of energies in Fig. \ref{figure4}. Overall, the
strengths of both real and imaginary parts of scalar and vector
potentials are sensitive to the asymmetry parameter in nuclear matter. The
energy dependence is visibly influenced by the asymmetries for all
Dirac potentials below energy $E=$100 MeV. The imaginary parts of the
scalar and vector potentials are much weaker than those of the real
parts. The behaviors of the imaginary parts are different from those
of the real parts and their energy dependence is obviously stronger.
The asymmetry dependence of the imaginary potentials for neutrons are
much smaller than those for protons near the Fermi surface, but
comparable at higher energies.

The optical potentials of a nucleon scattering off finite nuclei are
obtained by means of the local density approximation (LDA)
as usual, in which the spacial function for RMOP at the incident
energy $E$ can directly be related to the density ($\rho$),
momentum (\textit{k}) and asymmetry parameter ($\beta$) of nuclear
matter by
\begin{eqnarray}
      U_{LDA}(r,E)=U_{NM}(k,\rho(r),\beta), \label{eq9}
\end{eqnarray}
where $U_{LDA}$ and $U_{NM}$ correspond to the potentials of
a finite nucleus and the nuclear matter, respectively,
$\rho(r)$ is the density at nuclear radius \textit{r} and
related to the Fermi momentum $k_F$ by $\rho=(2k_F^3)/(3\pi^2)$.
  To obtain
the scattering amplitude and evaluate the observables for finite nuclei,
a Schr\"{o}dinger type equation is obtained by eliminating the lower
components of the Dirac spinor in a standard way. The equation for
the upper components of the wave function is transformed into

\begin{eqnarray}
\left[-\frac{\nabla^2}{2{\varepsilon}}+V^t_{cent}+V^t_{s.o.}(r)\vec{\sigma}\cdot\vec{\textbf{\L}}+V^t_{Darwin}(r)
\right]\varphi(\textbf{r}) \\
      \nonumber =\frac{\varepsilon^2-\textit{M}^2}{2\varepsilon}\varphi(\textbf{r}), \label{eq13}
\end{eqnarray}
where $V^t_{cent}$, $V^t_{s.o.}$ and $V^t_{Darwin}$ represent
the Schr\"{o}dinger equivalent central, spin-orbit
and Darwin potentials, respectively.
These potentials are related to the scalar $U_s$ and vector $U_0$ potentials by
\begin{eqnarray}
      \nonumber V^t_{cent}=\frac{M}{\varepsilon}U_s^t+U_0^t+\frac{1}{2\varepsilon}[U_s^{t2}-(U_0^t+V_c)^2], \label{eq14}
\end{eqnarray}
\begin{eqnarray}
      V_{s.o.}^t=-\frac{1}{2{\varepsilon}rD^t(r)}\frac{dD^t(r)}{dr}, \label{eq15}
\end{eqnarray}
\begin{eqnarray}
     \nonumber V_{Darwin}^t=\frac{3}{8{\varepsilon}D^t(r)}[\frac{dD^t(r)}{dr}]^2-\frac{1}{2{\varepsilon}rD^t(r)}\frac{dD^t}{dr} \\
     \nonumber -\frac{1}{4{\varepsilon}D^t(r)}\frac{d^2D^t(r)}{d^2r},\label{eq16}
\end{eqnarray}
where $V_c$ is the Coulomb potential for a charged particle, $D$
denotes a  quantity defined as
\begin{eqnarray}
       D^t(r)=M+\varepsilon+U_s^t(r)-U_0^t(r)-V_c. \label{eq17}
\end{eqnarray}
One can see again that the spin-orbit term, $V_{s.o.}$, emerges in a very natural
way from the relativistic scheme.

\begin{figure}[htbp]
\centerline{\includegraphics[width = 3.2in]{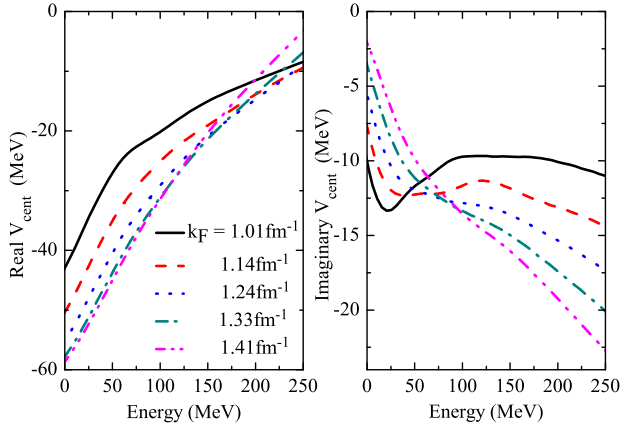}} \caption{
(color online)The Schr\"{o}dinger equivalent central potentials as
functions of the incident energies in symmetric nuclear matter at the
nuclear Fermi momenta $k_F = 1.01, 1.14, 1.24, 1.33,
1.41$ fm$^{-1}$. The real and imaginary parts of $V_{cent}$ are
plotted in left and right panels, respectively. } \label{figure5}
\end{figure}

Based on the Dirac potentials above, the Schr\"{o}dinger equivalent
potentials can be computed following Eqs. (\ref{eq15})-(\ref{eq17}).
Firstly, in Fig. \ref{figure5}, the central potential $V_{cent}$ in
the case of symmetric matter is shown as a function of incident energy at
various densities, corresponding to
$k_F$=1.01, 1.14, 1.24, 1.33, 1.41 fm$^{-1}$. For this figure we
dropped the Coulomb contribution to $V_{cent}$.
It is noted that the depths of real $V_{cent}$ decrease with
increasing energy, and there is a crossing of curves for different
densities around $E$ = 200 MeV.  This implies that the so-called
"wine-bottle bottom" shape appears at the the nuclear surface of
finite nuclei\cite{Ma88}. At energies below 50 MeV the
imaginary potentials $V_{cent}$ are deepest around $k_F$ = 1
fm$^{-1}$ and become weaker at higher densities. The shape of the
imaginary potentials for finite nuclei contains a local maximum
at the nuclear surface energies $E <$ 50 MeV, which indicates a
surface absorption. The situation is different at energies above 75 MeV,
where the absolute value of the imaginary potential increases
monotonically with the density, which exhibits the behavior of volume absorption.

\begin{figure}[htbp]
\centerline{\includegraphics[width = 3.2in]{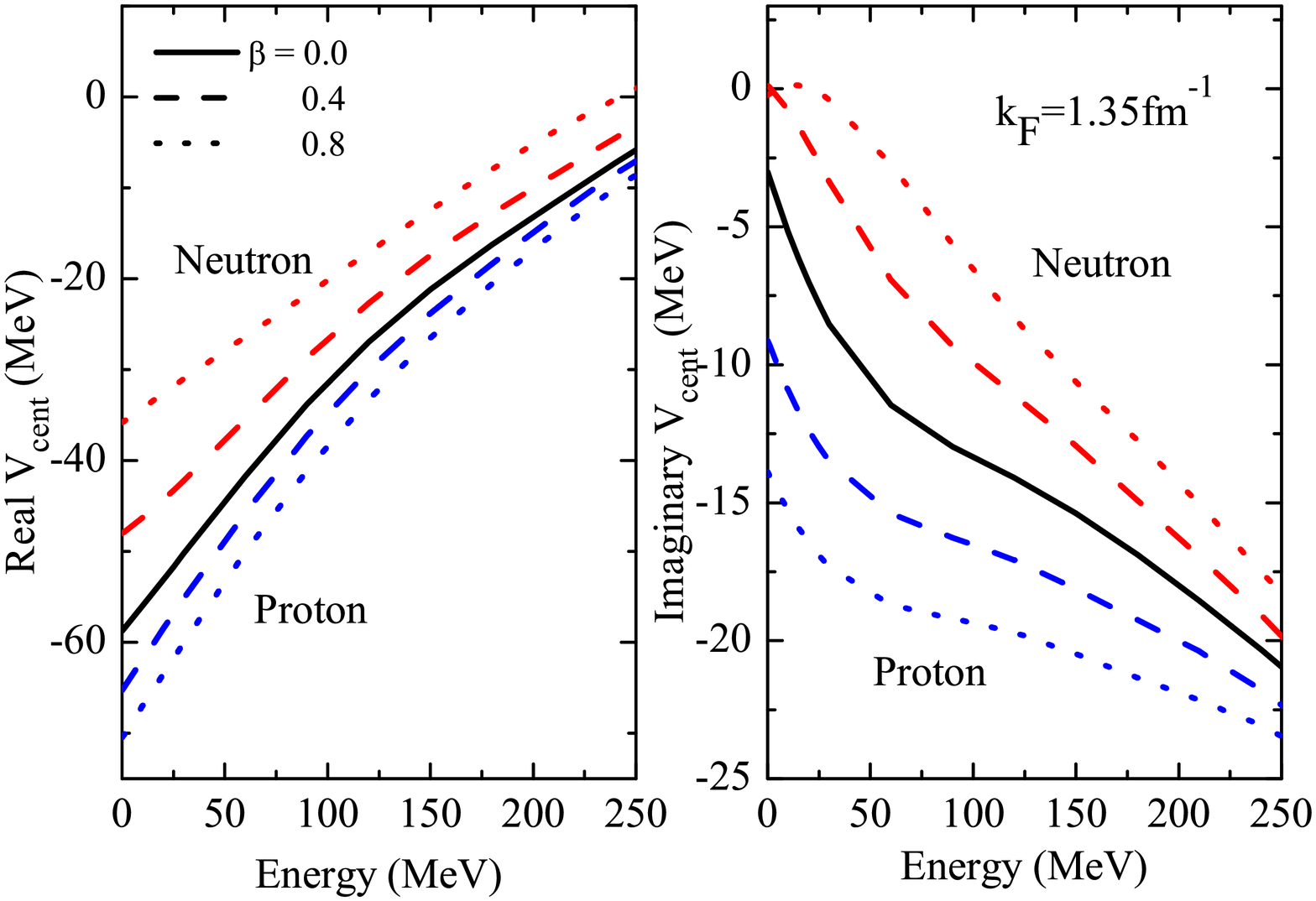}} \caption{(color
online) The Schr\"{o}dinger equivalent central potentials as
functions of the incident energies in the asymmetric nuclear matter at
$k_F=1.35$ fm$^{-1}$. The real part of $V_{cent}$ is in the left
panel. The right panel shows the imaginary part. The color
notations are the same as Fig. \ref{figure4}.} \label{figure6}
\end{figure}

\begin{figure}[htbp]
\centerline{\includegraphics[width = 3.2in]{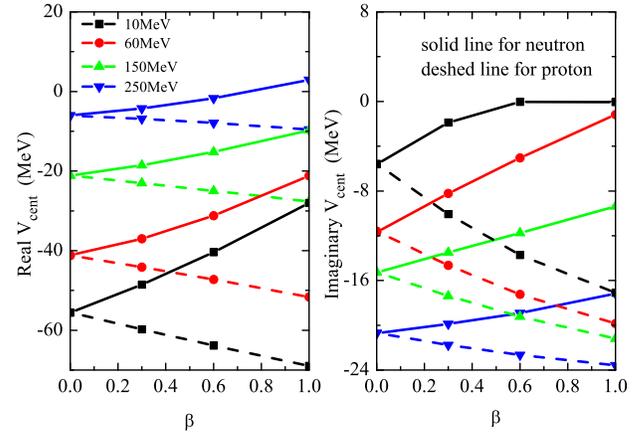}} \caption{(Color
online) The central potentials as functions of asymmetry $\beta$ at
$k_F=1.35$ fm$^{-1}$. The black, red, green and blue ones correspond
to the incident energies $E$ = 10, 60, 150, 250 MeV, respectively.
Solid curves are for neutron and dashed ones for proton.}
\label{figure7}
\end{figure}

Additionally we compare the $V_{cent}$ of proton and neutron versus
the  projectile energy $E$ for various $\beta$ values at the fixed
density $k_F$ = 1.35 fm$^{-1}$ in Fig. \ref{figure6}. One finds
that the depths of both real and imaginary potentials for neutrons
are smaller than those for protons below $E$=250 MeV. We specially
analyze the dependence of $V_{cent}$ on the isospin asymmetry at several
energies in Fig. \ref{figure7}.

\begin{figure}[htbp]
\centerline{\includegraphics[width = 3.2in]{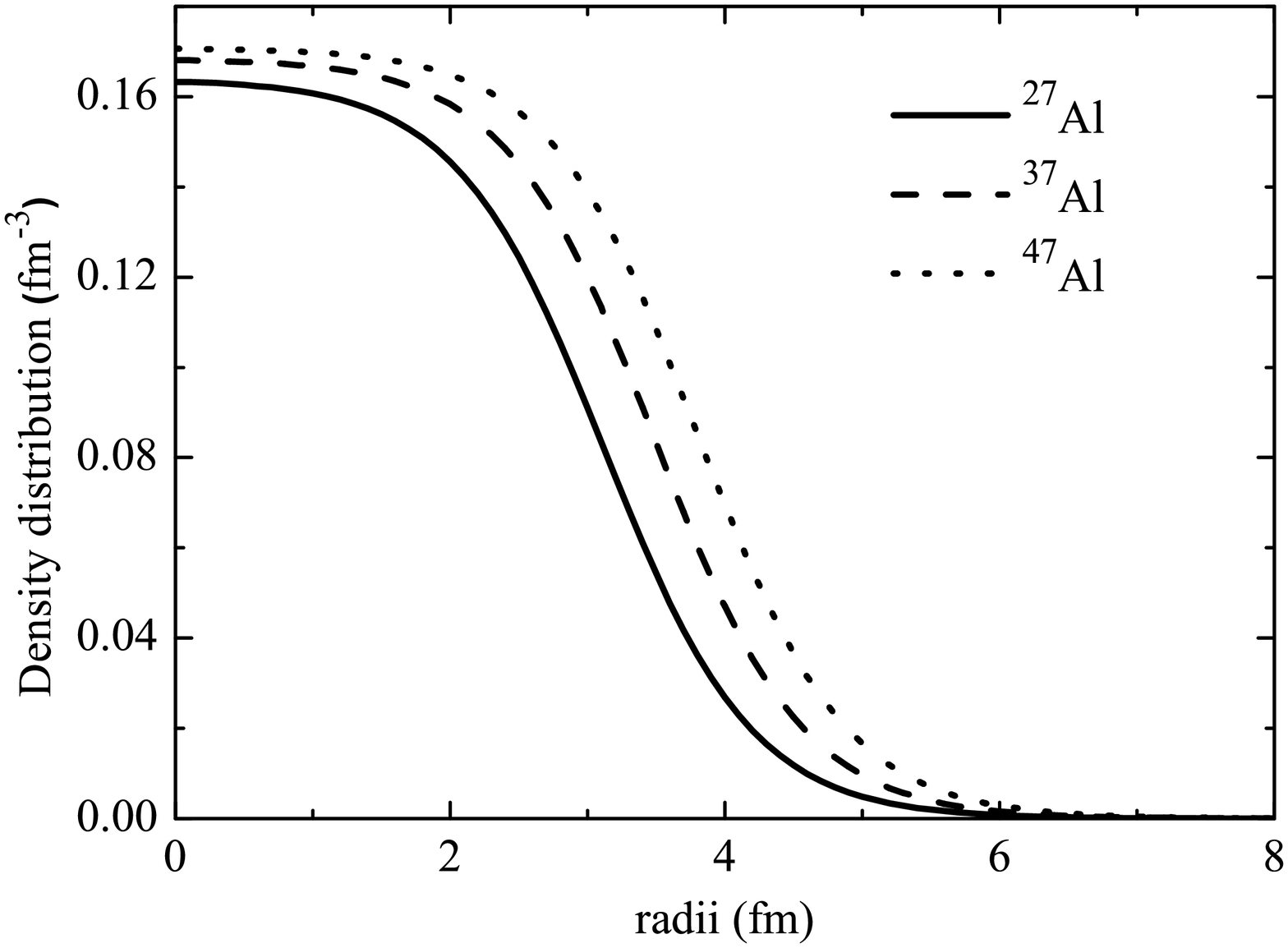}}
\caption{Nucleon densities of $^{27}$Al and two unstable isotopes
$^{37, 47}$Al} \label{figure8}
\end{figure}

As an example for the applications of RMOP we study the experimental
scattering data for the nucleon scattering off aluminum targets.
An empirical formula by Negele \cite{Negele70} is employed to
generate the nucleon density distribution of the stable $^{27}$Al
and two unstable neutron-rich isotopes $^{37,47}$Al, as shown in
Fig. \ref{figure8}. The asymmetry is assumed constant in the
interior of nucleus and determined by $\beta=(N-Z)/A$, where $N,Z,A$
indicate the numbers of neutron, proton and total nucleons of a
finite nucleus. Thus, the asymmetries of $^{27,37,47}$Al are
respectively taken as 0.037, 0.297 and 0.447.

In this discussion, we adopt the finite range correction with
Gaussian form to make  modifications in the potentials via LDA
approach, namely improved local density approximation
(ILDA)\cite{Jeu77}. The potential $U_{ILDA}$ of the finite nucleus
is finally obtained through the following integral,
\begin{eqnarray}
       U_{ILDA}(r,E)&=&(t\sqrt{\pi})^{-3}\times \label{eq18}\\
     && \int{U_{LDA}(r',E)exp(-|\vec{r}-\vec{r'}|^2/t^2)d^{3}r'}~, \nonumber
\end{eqnarray}
where $t$ is the parameter, that represents the effective range of
$U_{LDA}$ at radius $r'$. This is the only adjusted parameter in
this calculation and is selected as a constant 1.4 fm in all cases.

\begin{figure*}[htbp]
\centerline{\includegraphics[width = 5in]{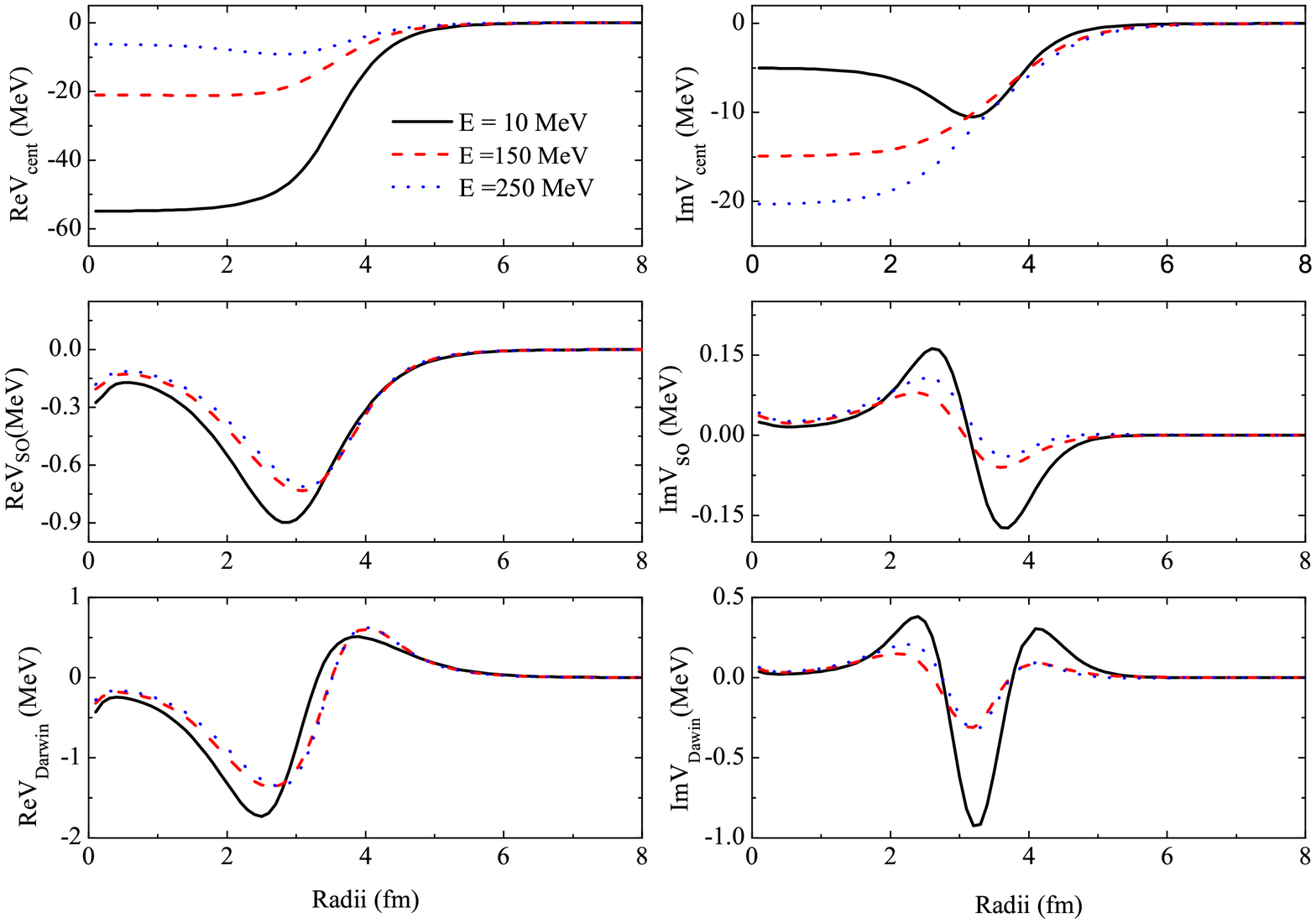}} \caption{(Color
online) The Schr\"{o}dinger equivalent central, spin-orbit and
Darwin potentials as functions of radius for $n+^{27}$Al at $E$ =10,
150, 250 MeV.} \label{figure9}
\end{figure*}

\begin{figure}[htbp]
\centerline{\includegraphics[width = 3.2in]{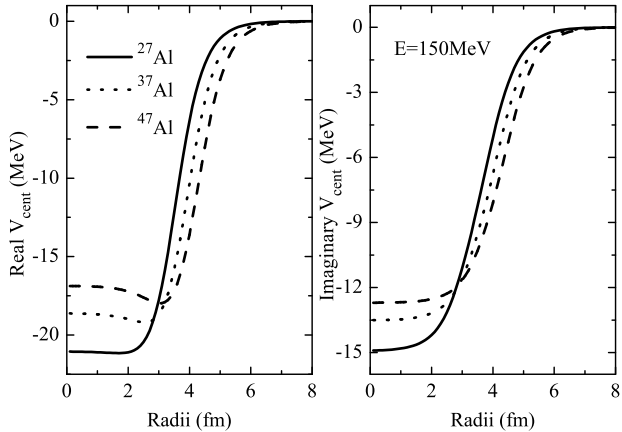}} \caption{The
$V_{cent}$ of $n+^{27,37,47}$Al at $E$ = 150 MeV. The real parts
are shown in the left and the imaginary ones in the right.}
\label{figure10}
\end{figure}

\begin{figure}[htbp]
\centerline{\includegraphics[width = 3.2in]{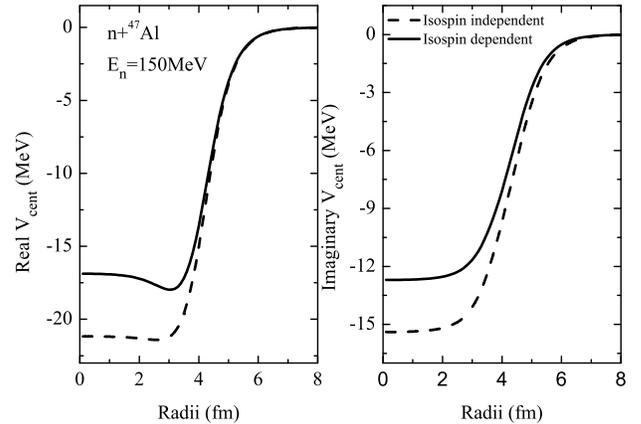}} \caption{The
$V_{cent}$ of $n+^{47}$Al in the isospin dependent mode and those
for neglecting the isospin asymmetry at $E$ = 150 MeV. The left panel
shows the real parts, the right one for the imaginary parts.}
\label{figure11}
\end{figure}

We derive the Schr\"{o}dinger equivalent potentials, $V^t_{cent}$,
$V^t_{s.o.}$, $V^t_{Darwin}$, for $n, p+^{27,37,47}$Al through this
ILDA approach. The potentials of neutron scattered from the stable
$^{27}$Al at $E$ = 10, 150, 250 MeV are plotted in Fig.
\ref{figure9}. The depths of the real  potentials decrease as the
energy increases and the imaginary parts change from the surface
absorption to the volume absorption. These behaviors are
consistent with the phenomenological OMP. It can also be observed
that the main contribution of RMOP comes from the central potential
$V_{cent}$, the spin-orbit $V_{s.o.}$ and Darwin $V_{Darwin}$ terms
are the necessary supplements to the central potential. In Fig.
\ref{figure10}, we construct the $V_{cent}$, for neutron scattered
from $^{27}$Al and the neutron-rich $^{37, 47}$Al. Due to the
varying nucleon density distributions and asymmetries, the depths of
$V_{cent}$ of the Al isotopes become smaller in the interior of the
nuclei ($r<3$ fm) with the increasing neutron numbers, while the
potentials of heavier nuclei are expanded for a large radius.
Meanwhile, we analyze $V_{cent}$ of $n + ^{47}$Al at $E=150$ MeV in
Fig. \ref{figure11} to check the impact of isospin asymmetry on the
potential. It is found that the central potentials of $^{47}$Al are
clearly affected by the isospin inside the nucleus. The impact is
believed to become stronger for those nuclei with larger values for $\beta$.


\section{RESULTS AND DISCUSSION}
\begin{figure}[htbp]
\centerline{\includegraphics[width = 3.2in]{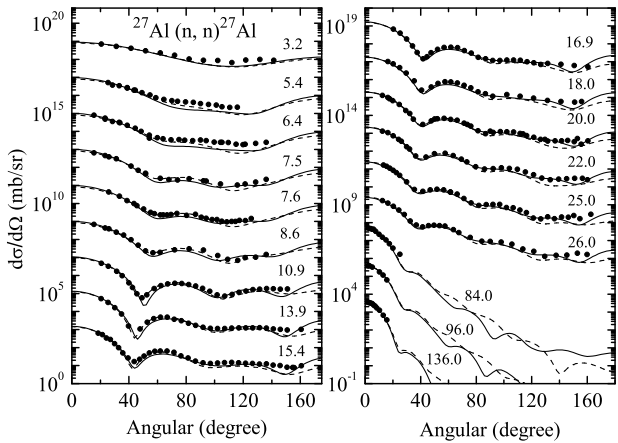}} \caption{
Angular distributions of $n+^{27}$Al elastic scattering at 3.2 MeV
$<E<$ 136 MeV. The curves and data points at the bottom represent
their true values, the others are multiplied by factors of 10$^2$,
10$^4$, etc. The solid curves and dashed ones indicate the results
calculated with RMOP and KD potentials, respectively. The dots are
experimental data listed in Table \ref{tab1}.} \label{figure12}
\end{figure}
\begin{figure}[htbp]
\centerline{\includegraphics[width = 3.2in]{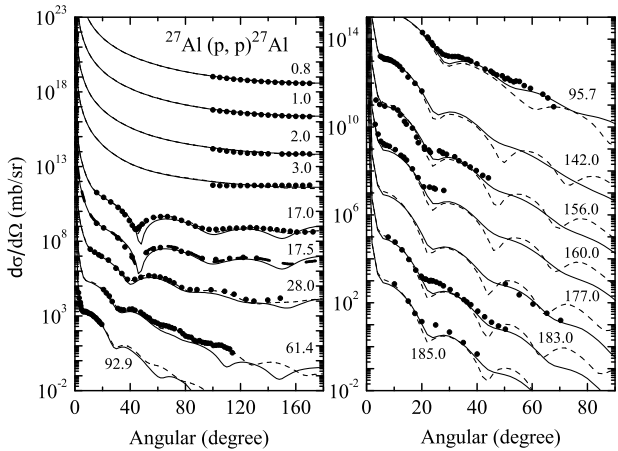}} \caption{Same
as in Fig.\ref{figure12}, except for $p+^{27}$Al elastic scattering
for induced energy 800keV$<E<$185.0 MeV.  The experimental data are
shown in Table \ref{tab2}.} \label{figure13}
\end{figure}
\begin{figure}[htbp]
\centerline{\includegraphics[width = 3.2in]{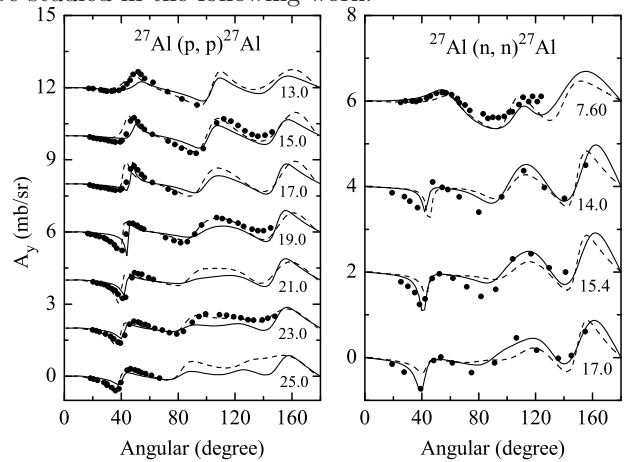}}
\caption{Analyzing powers of $p, n+^{27}$Al elastic scattering. The
curves and data points at the bottom represent their true values,
the others are shifted by increments of 2, 4, etc. The notations are
the same as in Fig. \ref{figure12}. The experimental data are listed
in Table \ref{tab1}-\ref{tab2}.} \label{figure14}
\end{figure}
\begin{figure}[htbp]
\centerline{\includegraphics[width = 3.2in]{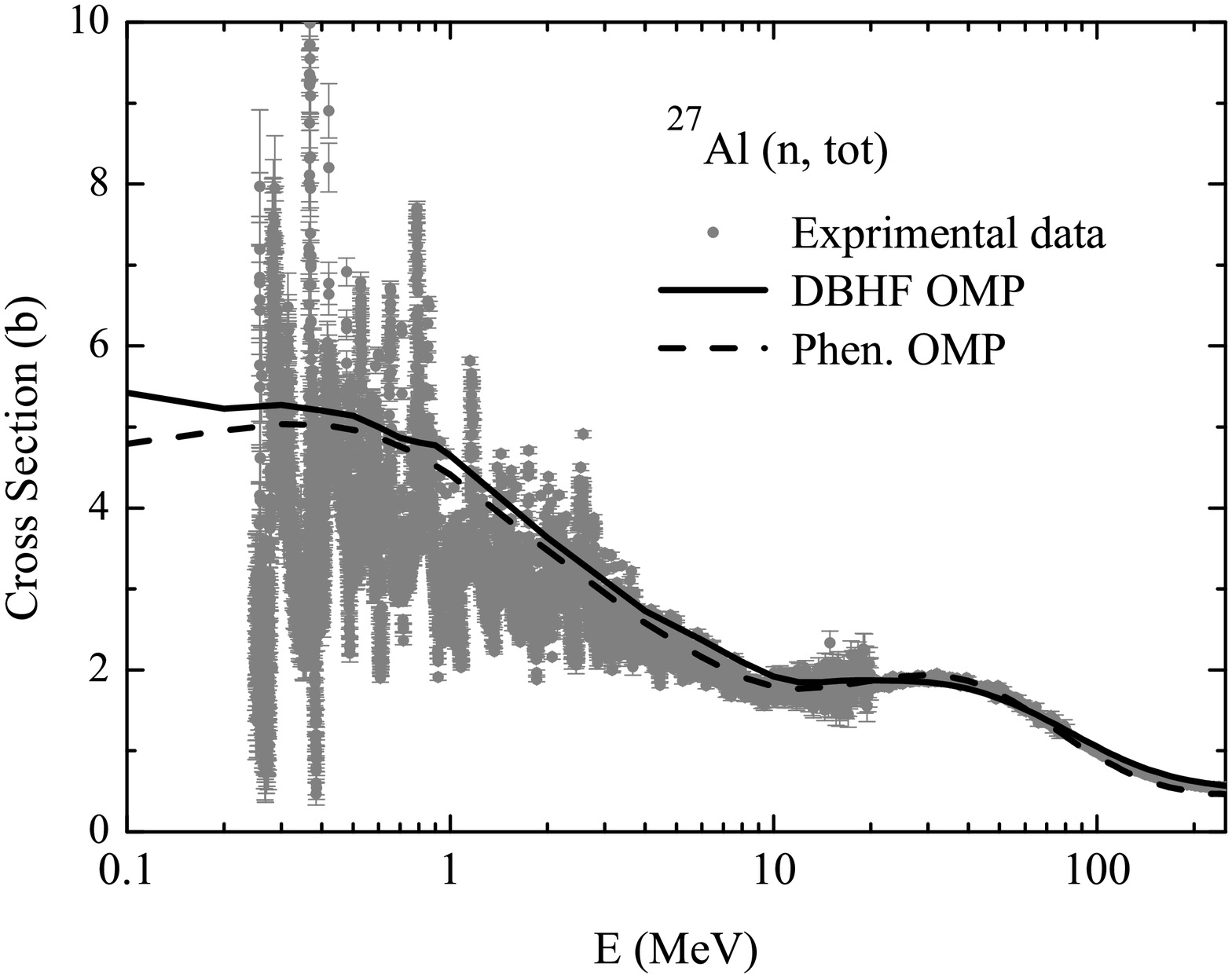}} \caption{The
cross sections of $n +^{27}$Al(n,tot). The experimental data are
listed in Table \ref{tab1}. The Phenomenological OMP indicates the
results calculated with the KD potentials.} \label{figure15}
\end{figure}

In this section, we discuss the angular distributions and
analyzing powers of $n, p+ ^{27}$Al and the cross sections of
$^{27}$Al($n$,tot) at 100 keV$<E<$250 MeV. At very low energies
the elastic scattering from the compound nucleus is sizable in
addition to those from the OMP. In order to compare with the
experimental data the Hauser-Feshbach model is utilized to determine
the contribution from the compound nucleus. For a particular
incident energy, there are six open single particle emission
channels from the compound nucleus, including the neutron, proton,
deuteron, tritium, $\alpha$, Helium-3 channels. The first several
discrete states are taken as the competing channels. For higher
excitation energies, a continuum described by the Gilbert¨CCameron
level density formula\cite{Gilbert65} is provided as
compensation. Therefore, the computed observables contain the
contributions from both shape and compound elastic scattering
processes. All calculations mentioned above are performed with the
assistance of the APMN code\cite{Shen02}.

The present calculations are systematically compared with the
experimental data and the results obtained with the global phenomenological
OMP, Koning-Delaroche (KD) potential \cite{Kon03} in Figs.
\ref{figure12}-\ref{figure15}. The experimental data in these
figures, the total cross sections $\sigma_{tot}$, the differential
elastic scattering cross sections $d\sigma/d\Omega$ and the
analyzing powers $A_y(\theta)$, are assembled in Table \ref{tab1} -
\ref{tab2}. The calculated observables of the neutron and proton
induced reactions are both in  good agreement with experimental
data. Moreover, the results of the RMOP are very similar to those
with the global KD potential. In some energy regions our predictions
are even better than those with the phenomenological ones, such as
the angular distributions of $p+^{27}$Al at $E>95.7$ MeV in Fig.
\ref{figure13}. As a conclusion, it is proved that the present RMOP
based on the DBHF approach is suitable to describe the potentials of
$n, p+^{27}$Al, especially for the spin-orbit term $V_{s.o.}$, which
is validated through the apparent agreement to the measured analyzing
power data in Fig. \ref{figure14}.

\begin{figure}[htbp]
\centerline{\includegraphics[width = 3.2in]{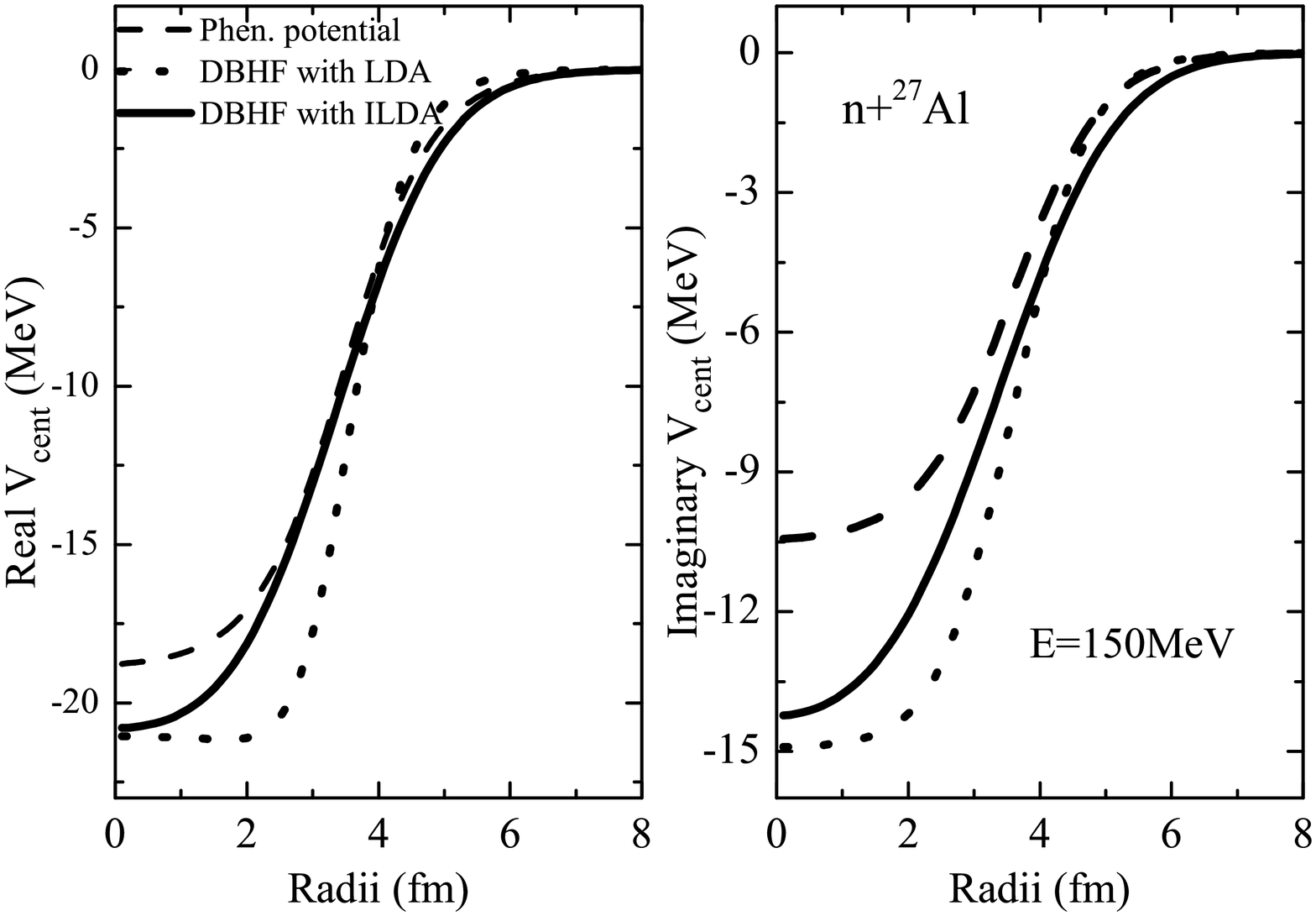}} \caption{ The
center potentials of $n +^{27}$Al at 150 MeV within this DBHF
approach in LDA and ILDA as well as the phenomenological approach
(labeled as Phen. potential in this figure).} \label{figure16}
\end{figure}

We particularly investigate the DBHF central potential of
$n+^{27}$Al at $E=150$ MeV with the ILDA approach and the
conventional LDA approach in Fig. \ref{figure16}. The $V_{cent}$ of
the phenomenological KD model is also displayed in this figure. It is
observed that the depths of the real and imaginary $V_{cent}$ in LDA
method are obviously deeper and exhibit a smaller range than those of the KD
potentials. This differences are reduced after the finite range
corrections in ILDA approach are taken into account,
and the corresponding RMOP looks
closer to KD ones. Therefore, rather similar scattering results in Figs.
\ref{figure12}-\ref{figure15} are obtained with both potentials.

\begin{table}
\caption{\label{tab1} $\sigma_{tot}$, $d\sigma/d\Omega$ and $A_y(\theta)$ database for $n+^{27}$Al}
\begin{ruledtabular}
\begin{tabular}{ccc}
Data &Ref.&Energy(MeV) \\
\hline
$\sigma_{tot}$ &\cite{ROHR94} & 0.25 - 27.0\\
&\cite{FINLAY93} & 5.3 - 250.0\\
&\cite{LARSON81} & 2.0 - 80.1\\
$d\sigma/d\Omega$ &\cite{Becker66} & 3.2  \\
                  &\cite{KINNEY70} & 5.4, 6.4, 7.5, 8.6 \\
                  &\cite{DAGGE89} & 7.6  \\
                  &\cite{WHISNANT84}& 10.9, 13.9, 16.9 \\
                  &\cite{Nagadi03} & 15.4 \\
                  &\cite{PETLER85} & 18.0, 20.0, 22.0, 25.0, 26.0\\
                  &\cite{Bratenahl50} & 84.0\\
                  &\cite{Salmon60} & 96.0 \\
                  &\cite{ZYL56} & 136.0  \\
$A_y(\theta)$     &\cite{DAGGE89} & 7.6 \\
                  &\cite{MARTIN86} & 14.0, 17.0\\
                  &\cite{Nagadi03} & 15.4\\
\end{tabular}
\end{ruledtabular}
\end{table}

\begin{table}
\caption{\label{tab2} $d\sigma/d\Omega$ and $A_y(\theta)$ database for $p+^{27}$Al}
\begin{ruledtabular}
\begin{tabular}{ccc}
Data&Ref.&Energy(MeV)\\
\hline
$d\sigma/d\Omega$ &\cite{Chiari01} & 0.8, 1.0, 2.0, 3.0 \\
                  & \cite{DAYTON56} & 17.0 \\
                  &\cite{CRAWLEY68} & 17.5 \\
                  &\cite{Dittman69} & 28.0\\
                  &\cite{FULMER69} & 61.4 \\
                  &\cite{Gerstein57} & 92.9, 95.7\\
                  &\cite{TAYLOR61} & 142.0 \\
                  &\cite{COMPARAT74} & 156.0\\
                  &\cite{JOHANSSON60,JOHANSSON61} & 160.0, 177.0, 183.0 \\
                  &\cite{Dahlgren67} & 185.0\\
$A_y(\theta)$     &\cite{Roy83} & 13.7, 15.0, 17.0, 19.0, 21.0, 23.0, 25.0\\
\end{tabular}
\end{ruledtabular}
\end{table}

\section{SUMMARY}
In this work, we study the RMOP for finite nuclei in the framework
of the DBHF approach with projection techniques using the Bonn-B bare $NN$ interactions.
Special attention is paid to the isospin dependence. The subtracted $T$-matrix representation is applied to minimize the
ambiguities, which arise due to the restriction to
positive energy states in the DBHF approach,  in the determination of the nucleon self-energies for
isospin asymmetric nuclear matter. In this way the precise form of the
nucleon self-energies with its momentum and isospin dependence are
determined. Thus, the nucleon RMOP is obtained from the results of these DBHF
calculations. An ILDA method
is adopted to intimately relate the density and isospin dependence
of the RMOP in nuclear matter to the radial dependence in finite
nuclei. A widely used empirical density distribution is employed for
finite nuclei. Only one free parameter  in ILDA is adapted in RMOP,
which is fixed to be $t$ = 1.4 fm for all calculations.

As an example of the applications we  investigate the elastic
scattering of $n, p+^{27}$Al at 100 keV $<E<$ 250 MeV. The predicted
observables in this approach are compared with experimental data and with
results from phenomenological KD potentials.  Good agreement is
 obtained. In addition, we perform a first analysis of the RMOP for
the unstable neutron-rich nuclei $^{37,47}$Al. It is demonstrated that the
isospin asymmetry has an obvious impact on the OMP potentials.

To apply this RMOP in a convenient way and make it accessible to other research
groups, explicit expressions for the nucleon scalar and vector potentials
might be desired. In this work, we systematically analyze the
energy, density and isospin dependence of the nucleon self-energies
and  the Dirac potentials in nuclear matter and the Schr\"{o}dinger
equivalent potentials in finite nuclei. It is observed that the
potential is sensitive to the isospin parameter $\beta$, but
some features related to the energy dependence are not sensitive to $\beta$.
 The full parameterization of this RMOP and the extensive application will
be studied in the following work.

\begin{acknowledgments}
This work has been supported by the National Natural Science Foundation of China(Grant Nos. 10875150 and 11175216);
the Deutsche Forschungsgemeinschaft (DFG) under contract no. Mu 705/5-2.
\end{acknowledgments}

\end{document}